\documentclass{article}
\usepackage{ijcai23}

\usepackage{times}
\usepackage{soul}
\usepackage{url}
\usepackage[hidelinks]{hyperref}
\usepackage[utf8]{inputenc}
\usepackage[small]{caption}
\usepackage{graphicx}
\usepackage{amsmath}
\usepackage{amsthm}
\usepackage{booktabs}
\usepackage{algorithm}
\usepackage{algorithmic}
\usepackage[switch]{lineno}

\title{Adaptive Attack Detection in Text Classification: Leveraging Space Exploration Features for Text Sentiment Classification}

\author{
Atefeh Mahdavi$^1$\and
Neda Keivandarian$^1$\and
Marco Carvalho$^{1}$
\affiliations
$^1$Florida Institute of Technology
\emails
amahdavi@fit.edu,
nkeivandaria2020@my.fit.edu,
mcarvalho@fit.edu
}

\begin{document}

\maketitle

\begin{abstract}
    Adversarial example detection plays a vital role in adaptive cyber defense, especially in the face of rapidly evolving attacks. In adaptive cyber defense, the nature and characteristics of attacks continuously change, making it crucial to have robust mechanisms in place to detect and counter these threats effectively. By incorporating adversarial example detection techniques, adaptive cyber defense systems can enhance their ability to identify and mitigate attacks that attempt to exploit vulnerabilities in machine learning models or other systems. Adversarial examples are inputs that are crafted by applying intentional perturbations to natural inputs that result in incorrect classification. In this paper, we propose a novel approach that leverages the power of BERT (Bidirectional Encoder Representations from Transformers) and introduces the concept of Space Exploration Features. We utilize the feature vectors obtained from the BERT model's output to capture a new representation of feature space to improve the density estimation method.
\end{abstract}

\section{Introduction}
In the context of adaptive cyber defense, the detection of adversarial examples provides several benefits. As attacks evolve and new types of adversarial examples emerge, the system can leverage detection algorithms to analyze incoming data streams and update its defense accordingly. This adaptive approach ensures that the defense mechanisms stay effective against the evolving threat landscape. Furthermore, by studying the characteristics and patterns of detected adversarial examples, adaptive cyber defense systems can gain insights into the strategies and techniques employed by attackers. This knowledge can be used to refine defense strategies, identify vulnerabilities, and proactively strengthen the system's resilience against further attacks.\par

This study aims to contribute to the field of text classification by focusing on adversarial example detection in text sentiment classification. By improving the detection and understanding of adversarial examples in sentiment classification, we aim to advance the effectiveness of adaptive cyber defense mechanisms and ensure the reliability of sentiment-based applications in dynamic and ever-changing environments.\par

Sentiment Classification is a task in natural language processing (NLP) that involves determining the sentiment polarity of a given piece of text, such as reviews in various applications, including customer feedback analysis, social media monitoring, and market research. The goal is to classify the text as positive, negative, or neutral based on the sentiment it conveys \cite{keivandarian2023survey}. In the context of cyber attacks, sentiment analysis can help organizations to detect threats and stay ahead of evolving cyber risks and protect critical assets and sensitive information effectively. By analyzing textual data, organizations can gain insights into sentiments expressed in online discussions, news articles, and social media, enabling them to detect and prevent potential cyber threats. This technique aids in early threat detection by identifying emerging risks through sentiment analysis of discussions related to vulnerabilities or targeted organizations. Moreover, sentiment analysis helps monitor hacker forums and the dark web, providing valuable information on threat actor motivations and potential targets. It can also assist in identifying insider threats by analyzing sentiments expressed in employee communications. During cyber incidents, sentiment analysis helps gauge public sentiment and perception, allowing organizations to manage incidents effectively. \par
\begin{figure*}[h]
    \centering
    \includegraphics[width=0.9\textwidth]{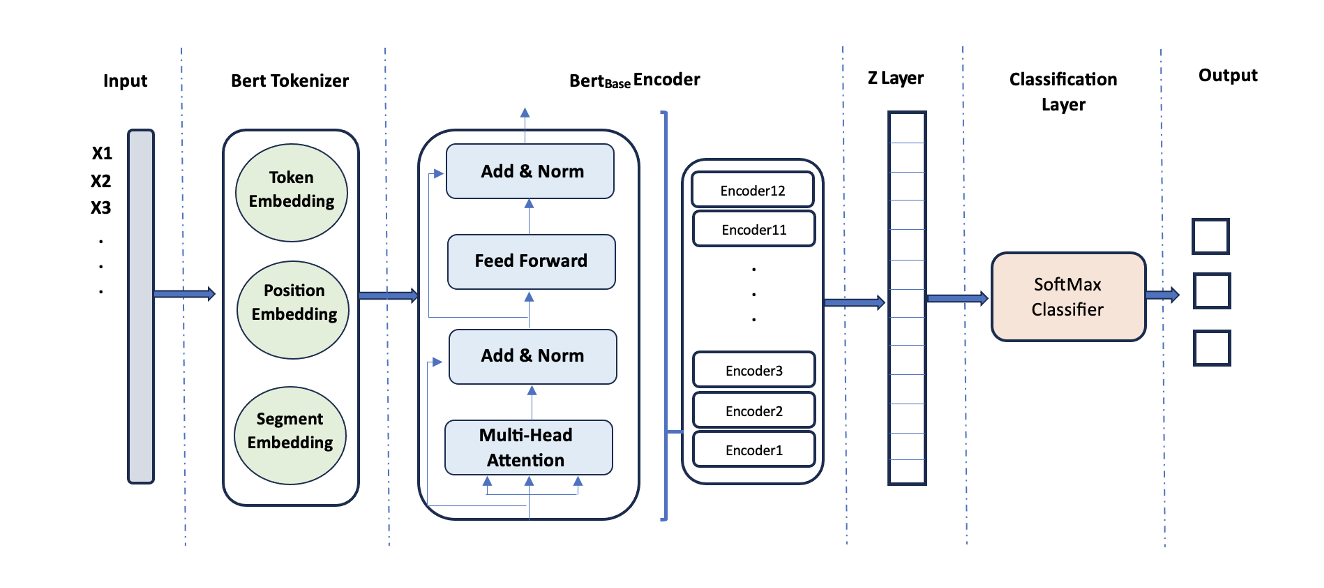}
    \vspace{-0.8em}
    \caption{Process of Fine-tune BERT model by transforming the feature space (Z Layer) and enhancing class separation. Using the final classification layer, the BERT model predicts and detects adversarial examples.}
    \label{fig4}
\end{figure*}

With the rapid advancement in the area of machine learning, DNNs have emerged as an efficient and accurate framework that can be applied to various core security problems such as malware detection. However, recent studies have demonstrated that the accuracy of these models can be reduced facing adversarial examples \cite{szegedy2013intriguing}. These inputs are derived from natural inputs in the training set and testing set. Suppose $F$ is a trained DNN classification model and $x$ is a natural input that is correctly classified, i.e., $C(x)=l$, where $C(x)$ means the classification of $F$ on $x$ predicted as class $l$. Then, an adversary can produce a new input ${x^*}$ that is similar to $x$ but is classified incorrectly, i.e.,  $C({x^*})\neq l$. In this simple case, the input is misclassified in a class different from the legitimate source class. A more restrictive case is where the adversary chooses a specific target class $t \neq \l$ and crafts the adversarial example ${x^*}$ close to $x$, and yet that gets misclassified such that $C({x^*})=t$. Adversarial example generation methods exploit gradient-based optimization for normal samples to find a small perturbation in a direction that maximizes the chance of misclassification.

The radius of the search area around the natural samples and the used loss function to guide the direction are parameters that differ in these techniques.\\
Multiple strategies have been proposed for adversarial attacks on neural network models \cite{carlini2017towards}, along with ongoing efforts to  understand adversarial samples to improve the robustness of neural networks against them \cite{gu2014towards}. On the detection of adversarial samples, \cite{papernot2016distillation} proposed defensive distillation technique as a way of preventing the network from fitting too tightly to the data. In this technique, the Softmax outputs of the neural network that was trained on the training data are served as the input to train another DNN. \cite{metzen2017detecting} uses the inner convolutional layers of the network to detect adversarial examples.\par

Transformers-based models have significantly advanced sentiment classification in NLP tasks. The introduction of models like BERT (Bidirectional Encoder Representations from Transformers) has revolutionized the field by achieving state-of-the-art results on various sentiment analysis benchmarks\cite{devlin2019bert}, \cite{sun2020adv}.
One of the key advantages of transformers in sentiment classification is their ability to capture contextual information effectively. Unlike traditional approaches that rely on fixed-length features or bag-of-words representations, transformers consider the entire context of a sentence or document.\par
BERT is an architecture that consists of a stack of transformer encoder layers. Each layer in BERT comprises several sub-layers, including a multi-head self-attention mechanism that allows the model to attend to different parts of input sequence and capture dependency between tokens and position-wise feed-forward neural networks that apply non-linear transformation to the taken representation.\par

To use BERT for text classification, the input text data is preprocessed, tokenized, and encoded into numerical representations using the BERT tokenizer and vocabulary. After padding or truncating the sequences to a fixed length, a pre-trained BERT model, BERT-base with 12 encoder layers, is employed. The model, consisting of transformer encoder layers, is fine-tuned on task-specific labeled data by adding a classification layer and optimizing a loss function.\par To further enhance the performance of text classification and mitigate adversarial attacks, we propose an additional layer that can be added before the final classification layer in the fine-tuned BERT model. This additional layer can serve as a defense mechanism against adversarial examples by introducing robustness and improving the model's ability to distinguish between genuine and manipulated inputs. Figure 1 shows the configuration of the proposed idea. During the fine-tuning process, the BERT model first learns to transform the feature space into a new space where the classes become further apart from each other. By passing these representations through the final classification layer, the BERT model can make predictions on new, unseen text data, assigning it to appropriate class labels based on its understanding of the learned task-specific patterns. This allows the model to capture task-specific patterns and improve its ability to classify text accurately. \par
The proposed fine-tuned BERT model represents a feature space and calculates density estimation. This idea can be an effective defence method against adversarial examples. In Section 2, we explain density estimation strategy to solve this problem. In Section 3, we introduce our proposed framework and conclude the position paper in Section 4.

\section{Detecting Adversarial Samples by Performing Density Estimation}
\begin{figure*}[h]
    \centering
    \includegraphics[width=13cm]{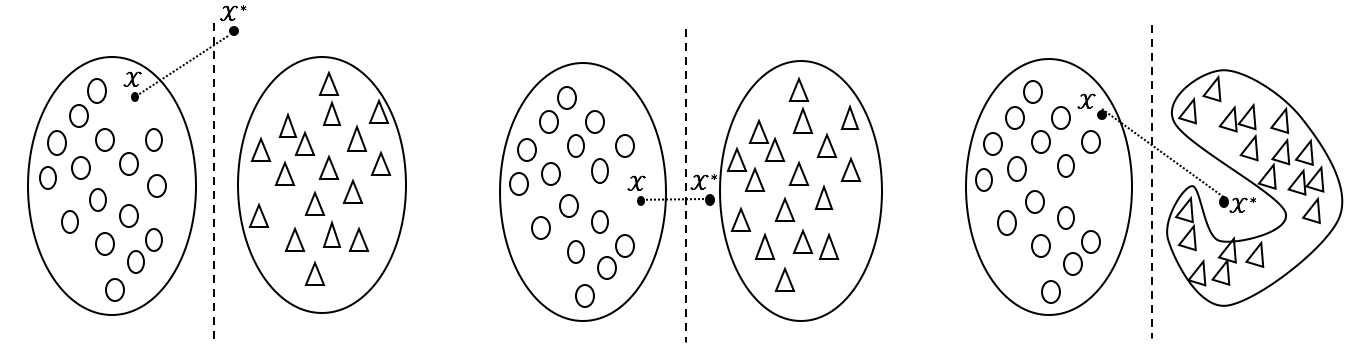}
    \vspace{-0.8em}
    \caption{(a): The adversarial example $x^{*}$ is generated by traversing away from ‘$\circ$’ sub-manifold, but it is still far away from the ‘$\bigtriangleup$’ sub-manifold. (b): the adversarial example ${x^*}$ is near the ‘$\bigtriangleup$’ sub-manifold but not on it, and it is also near the decision boundary (dashed line). (c): there exists a pocket in ‘$\bigtriangleup$’ sub-manifold that allows ${x^*}$ lying in sub-manifold of the wrong label and far away from classification boundary.}
    \label{fig4}
\end{figure*}
~\cite{tanay2016boundary} proposed the ‘boundary tilting’ perspective to explain the existence of adversarial samples for DNNs. They argued that adversarial examples stay near the classification boundaries and in regions where these boundaries are close to a sub-manifold of the data. ~\cite{gardner2015deep} proposed that the true label of natural inputs can be changed by moving away from the manifold of the training data. Based on these studies, ~\cite{feinman2017detecting} assumed that by moving away from a point $x$ belonging to a source class $C(x)=l$, the generated adversarial sample ${x^*}$ must be pushed off of the data manifold and is classified incorrectly as $C({x^*})=\acute{l}$. Figure 2 depicts two sub-manifolds of circle and triangle that are separated by classification boundary (dashed line). Three possible situations for the location of adversarial examples are shown in a, b, and c insets, respectively. In these two-dimensional binary classification settings, ${x^*}$ lies off its associated manifold.

This work is based on performing kernel density estimation on the outputs of the final hidden layer learned by the model. The motivation behind using the outputs of the final layer is that this layer can provide more simplified manifolds to work with than input space. Then, they investigate how far point $x$ is from the final predicted class $l$. They argued that adversarial samples are likely to be in region of lower density estimates. Therefore, the classifier thresholds on the kernel density of the sample based on a selected threshold $\tau$ and reports $x$ as adversarial if its corresponding kernel density is less than $\tau$. 

While this approach can easily detect an adversarial example that is far from ‘$\bigtriangleup$’ sub-manifold (Figure 2.a), it does not work well for two cases: when $x^*$ falls near ‘$\bigtriangleup$’ sub-manifold (Figure 2.b), and for a more difficult detection when $x^*$ lies in the pocket of wrong sub-manifold (Figure 2.c). They showed that the density estimation of the adversarial sample decreases for the correct class and increases for the incorrect class. In this paper, we expand an alternative approach that can cover all cases.

\section{Approach}
BERT can provide a highly informative latent representation of input data, particularly in the context of natural language processing tasks. This learned latent representation tends to yield excellent performance in various applications. In the context of adversarial detection, BERT can be utilized to analyze and understand the textual content of input data, such as natural language text or sequential data. By feeding the input data through BERT, it can generate high-quality contextualized representations, often referred to as embeddings or encodings. These representations encode rich information about the input data, capturing its semantic meaning, context, and syntactic structure. The BERT-based representations can then be utilized as features for adversarial detection algorithms or classifiers. By comparing the BERT embeddings of input samples, it becomes possible to identify patterns or discrepancies that indicate the presence of adversarial examples. This study focuses on the representation outputs generated by BERT, and we propose a novel approach to enhance the effectiveness of instance representation. The motivation behind the proposed method is to discover improved representations for detecting adversarial examples. This is achieved by promoting greater separation between samples from different classes. Combining this new representation of feature space with kernel density estimation facilitates adversarial example detection. ~\cite{feinman2017detecting} claimed that the density estimation strategy failed to detect adversarial examples when they fall near the incorrect class (Figure 2.b) or in the pocket (Figure 2.c). In this new setting, while adversarial examples leaving the correct class and moving towards the incorrect class, they still stay far from the incorrect class. It helps to decrease the density estimate for adversarial examples; this, in turn, causes these examples to remain in a region of lower density estimates regardless of the original location of adversarial example $x^*$.

Specifically, given the point $x$ with predicted class $l$, the density estimate is defined as:
\begin{align}
  \hat{K}(x) = \ \frac{1}{\left | X_{l} \right |}\sum_{x_{i}\in X_{l}} exp(\frac{\left | Z(x)-Z(x_{i}) \right |^{2}}{\sigma^{2}})  
\end{align}%
Where $X_{l}$ is a set of training points with label $l$ and $Z(x)$ is the output of the last hidden layer of a neural network for point $x$.

The proposed representation of samples is inspired by a Fisher Discriminant. A Fisher Discriminant aims to find a linear projection that maximizes the separation of samples from different classes and minimizes the variance of the representation of the samples within a class. Such a representation is gained by optimizing Fisher's criteria. In this work, we aim to have a representation that pushes between-class distances and separates instances from different classes. In this setting different sub-manifolds are further apart and well separated which leads to larger spaces among them. Consequently, it is likely that the detection of adversarial examples can be achieved more easily.\\  

To satisfy this desirable property of the feature space, we define a second loss (Equation 2). It is a common approach that other studies have employed to train neural networks using more than one loss function. For example, the encoder network of an Adversarial autoencoder minimizes the reconstruction loss and the generators loss ~\cite{makhzani2015adversarial}. The distance measurement depicted in Equation 2 is computed using the mean of feature vectors extracted from BERT. 
\begin{align}
  \begin{matrix}Second\; Loss=
-(min \left \| \mu _{i}- \mu _{j}\right \|_{2}^{2})\\ 
\;\;\;\;\;\;\;1\leq i \leq N\\ 
\;\;\;\;\;\; i+1\leq j \leq N
\end{matrix} 
\end{align}%
In the above equation, $N$ is the number of all classes and $\mu _{i}$ is the mean of class $C_{i}$ and defined as:
\begin{align}
\mu _{i} = \frac{1}{N_{i}}\sum_{n\in C_{i}}^{} Z_{n}
\end{align}%
Here, $N_{i} $ is the number of training examples in class $C_{i}$ and $Z_{n}$ is feature vector output of instance $n$. The class means are used as a measure of separation. Therefore, the between-class separation is maximized in terms of the distance between the two closest class means ($\mu _{i}$ and $\mu _{j}$) among all the $N$ classes.\par
Typically, most of the DNNs use the Softmax function as the activation function in the output layer. 
The network is trained in the backward propagation by adjusting the value of weighted connections. The objective of the training is to minimize the loss function, such as cross-entropy that represents the difference between the output of the softmax function and the desired output to achieve a low classification error in the training data. This process leads to a set of properly adjusted weights that enables the neural network to be used effectively for the purpose it is initially designed for. Throughout this investigation, the neural network is trained on the proposed loss in Equation 2 along with the cross-entropy loss simultaneously.\par
We plan to train a DNN using mini-batch stochastic gradient descent with backpropagation. For evaluation purpose, we use Fast Gradient Sign Method (FGSM)\cite{goodfellow2014explaining} to generate adversarial samples and evaluate the performance of the proposed defense for two malware datasets: Android Malware dataset and MalwareTextDB.
The adversarial sample $x^*$ in FGSM method is calculated as
\begin{align}
    x^* = x + \epsilon \, sign(\bigtriangledown _{x}J(\theta ,x,y))
\end{align}
In this respect, $J(\theta ,x,y)$ denotes the model's loss function that specifies the cost of classifying the point $x$ as label $y$. $\theta$ represents the model parameters and $\epsilon$ is a binary constant that controls the perturbation magnitude. This attack uses the derivative of the loss function with respect to the input feature vector. The original input $x$ is perturbed in the direction of the loss gradient by magnitude $\epsilon$. A FGSM is a type of white-box attack in which the attacker has access to the loss function and model parameters, in contrast to black-box attacks that the intruder does not have knowledge of the model.   
\section{Conclusion}
In this paper we address the challenging task of detecting adversarial examples in text sentiment classification. In addition to advancing the field of text sentiment classification, our proposed approach holds significant potential for enhancing adaptive cyber defense systems. By leveraging the power of BERT and introducing Space Exploration Features, our novel approach can significantly enhance the robustness of sentiment analysis models, thereby fortifying adaptive cyber defense systems against malicious attacks that exploit vulnerabilities in natural language processing pipelines. The study leverages feature vectors derived from the BERT model's output to create an enhanced representation of the feature space, aiming to improve the accuracy of density estimation methods. The proposed approach  can detect adversarial examples in text sentiment classification, contributing to the advancement of robust and reliable models in this field. Further research should focus on implementing the proposed idea. We will conduct extensive experiments to evaluate  effectiveness and compare the performance of the proposed approach with existing methods for adversarial example detection in text sentiment classification.

\bibliographystyle{named}
\bibliography{main}

\end{document}